\documentclass[onecolumn,journal,a4paper]{IEEEtran}
\usepackage{times}
\usepackage{cite}
\usepackage{color}
\usepackage{epsf}
\usepackage{epsfig}
\usepackage{graphicx}
\usepackage{graphics}
\usepackage{epstopdf}
\usepackage[small]{caption}
\usepackage{amsmath}
\usepackage{amssymb}
\usepackage{amsthm}
\usepackage{amsxtra}
\usepackage[ruled]{algorithm2e}
\usepackage{algorithmic}
\usepackage{enumerate}
\usepackage{multirow}
\usepackage{enumerate}
\usepackage{amssymb}
\usepackage{lipsum}
\usepackage{setspace}
\usepackage{adjustbox}
\usepackage{multirow}
\usepackage{subcaption}
\usepackage{hyperref}
\usepackage{url}
\usepackage{color,soul}
\usepackage{geometry}
\usepackage[utf8]{inputenc}
\usepackage{array}
\usepackage{babel}

\usepackage{lipsum}
\usepackage{makecell}
\usepackage{draftwatermark}
\SetWatermarkText{Accepted in IEEE Industrial Electronics Magazine}
\SetWatermarkScale{0.1}

\usepackage[normalem]{ulem}

\usepackage{rotating} 
\usepackage{graphicx} 

\newcommand{\Rmnum}[1]{\expandafter\@slowromancap\romannumeral #1@}

\makeatother

\begin{document}

\title{Blockchain-enabled Circular Economy: Collaborative Responsibility in Solar
Panel Recycling}
\author{Mohammad Jabed Morshed Chowdhury, Naveed Ul Hassan, Wayes Tushar, Dustin Niyato, Tapan K. Saha, H. Vincent Poor, and Chau Yuen}

\maketitle

\onehalfspacing
\begin{abstract}
The adoption of renewable energy resources, such as solar power, is on the rise. However, the excessive installation of solar panels and a lack of recycling facilities pose environmental risks. This paper suggests a circular economy approach to address the issue. By implementing blockchain technology, the end-of-life (EOL) of solar panels can be tracked, and responsibilities can be assigned to relevant stakeholders. The degradation of panels can be monetized by tracking users' energy-related activities, and these funds can be used for future recycling. A new coin, the recycling coin (RC-coin), incentivizes solar panel recycling and utilizes decentralized finance to stabilize the coin price and supply issue. 
\end{abstract}

\section{Background \& Motivation}\label{Sec:Introduction}The $27^\text{th}$ Conference of the Parties (COP$27$) created a strong push for the world's sustainability. At Cop27, participating countries have reaffirmed their commitment to phase down all fossil fuels and increase ''low-emissions energy" \cite{Solar_Fiona_2022}. Subsequently, significantly more solar panels - along with other renewable energy resources - are expected to be installed in solar farms and households supplying a significant portion of the energy demand created by the discontinuation of coal power plants.

The existing and new installation of solar photovoltaics (PV) will scale up the generation and use of clean energy, positively contributing to limiting global warming to $1.5^\circ\mathrm{C}$. However, if not addressed urgently, these renewable assets could also adversely impact the climate and warm the atmosphere. This is because solar panels contain a layer of crystalline silicon sandwiched between the glass sheet and rigid polymer films, a toxic source of environmental pollutants~\cite{ToxicPanel_2018}. If these panels are thrown in landfills for disposal purposes, the toxic gas and dust resulting from this disposed silicon can significantly damage the environment~\cite{Neelam_WMR_2022} and human health~\cite{Matta_Silica_2017}. Further, when new panels replace these panels, the production of these panels produces harmful by-products like sulphuric acid, hydrogen fluoride, and nitric acid~\cite{Neelam_WMR_2022}, and also the procurement of raw materials involves systematic extraction of raw commodities from underdeveloped countries that involve environmental degradation, and where human rights violations have been criticized as unethical~\cite{Cobus_FP_June_2022}. Therefore, it is essential to not dispose of end-of-life (EOL) panels in landfills but instead, appropriately recycle the materials for reuse to minimize these impacts on the climate and human health. Otherwise, by 2050, the world will have an estimated total solar e-waste of 78 million tonnes~\cite{Conversation_2021} without any legitimate way to recycle it in an environmentally friendly and safe way.

Given this context, the concern for recycling renewable energy assets has gained momentum recently. A key challenge identified as a potential barrier to widespread recycling in the future is the considerable cost that recycling companies need to bear during the recycling process. The robust design of solar panels not only provides much-needed longevity but also makes it very hard to recover and liberate the materials from the panels when they reach their EOL. The US Department of Energy estimates that the recycling cost of a solar module (approximately 350 W rating) ranges from \$15 to \$45, compared with \$1 to \$5 for disposing of it in a landfill \cite{USDep_2021}. According to Toshiba Environmental Solutions~\cite{ToxicPanel_2018}, the high cost of recycling will make it challenging to create a profitable business, unless the recycling companies charge several times what is set by regulations. Therefore, innovations are needed in how we operate the overall renewable energy framework to balance expectations between recycling companies and government policymakers in ensuring that the cost of recycling is shared among all the beneficiaries of renewable energy and that the waste is correctly recycled to minimize its environmental impact.  

This paper proposes a new idea to proportionately shift the responsibility of recycling to the beneficiaries of renewable assets through a long-term planning solution. We argue that by using a modified concept of circular economy (CE)~\cite{Julian_RCR_Dec_2017} and taking advantage of emerging technology platforms like blockchain~\cite{Merlinda_RSER_Feb_2019}, individual users of solar energy can track their usage of assets, estimate the loss of lifetime per usage, and accordingly manage a monetary deposit that can help them to recycle their assets after the EOL. Thus, we introduce a new outlook on how various stakeholders can enjoy the benefits of renewable energy assets and take the responsibility of recycling them at the end of their lifetime by bearing the partial or complete cost of the process. Blockchain will act as a key enabler of our solution by keeping track of solar panels' EOL and energy production and ensuring the responsibilities of CE agents (i.e., users, manufacturers, utilities, and recyclers) are recorded in an immutable way and enforced through smart contracts. To the best of our knowledge, this is the first proposition of the idea of \emph{sharing responsibility} in solar panel recycling with a discussion of frameworks to achieve that. 

Certainly, numerous recent studies have posited that the incorporation of circular economy principles can significantly improve the sustainability of solar energy systems. This is achieved by promoting resource efficiency, minimizing waste, and prolonging the lifespan of components, as indicated by Farrell in 2020 \cite{Farrell_RSER_2020}. Our suggested circular economic model complements these advantages by facilitating the recycling of a greater number of solar panels, thereby aiding in the recovery of resources from EOL Panels. When solar panels reach the end of their lifespans, they can be disassembled, and their various components, such as silicon wafers, glass, and metals, can be either reused or recycled. This approach diminishes the demand for new materials and mitigates the environmental impact associated with the production of solar panels. The retrieval of valuable materials from EOL solar panels through recycling processes contributes to closing material loops and lessening the necessity for raw material extraction. Technologies for the efficient extraction and purification of materials from retired panels are currently under development, playing a role in establishing a more sustainable supply chain for solar energy systems.

We acknowledge that regulation also holds a pivotal role in shaping the processes of solar panel recycling within a country. Indeed, numerous nations, such as Australia \cite{PVAlliance}, are actively formulating regulations to address the management of solar assets at the end of their lifecycle. In this context, the exploration of potential technological solutions becomes a crucial factor in guiding governmental bodies in the development of these regulations. The recognition that recycling entails substantial costs underscores the importance of clear guidelines outlining which stakeholders will bear these responsibilities. The motivation behind the proposed work is to contribute a potential solution that aids in formulating such guidelines, thereby assisting the government in making well-informed regulatory decisions.

\section{Solar Recycling - Present Reality}\label{sec:SolarRecycle} The usual productive lifespan of solar panels is around $25$-$30$ years \cite{Otasowei_JETP_2015}. This relatively long lifetime makes owners, manufacturers, and suppliers rarely worry about the future upgrading of the panels during their installation. As a result, recycling solar panels has not been a concern for the first $25$ years of development \cite{Chowdhury_ESR_2020}. However, the surge in using solar panels in solar plants and on household rooftops began in $2010$ and increased at a very sharp rate in later years \cite{King_ABCNews_2021}. That means that, in years around 2030 to 2040, the world will have produced an extremely large amount of solar panel waste, which will only grow in the future. Based on the current practice, these high volumes of solar waste will end up in landfills at the EOL as solar panel waste still falls under the general waste classification in most countries from a regulatory perspective \cite{Attila_Blog_2021}. Thus, if the regulatory framework is not revised for solar panels, around $78$ million tonnes of solar waste will end up in landfills by $2050$~\cite{Chowdhury_ESR_2020}, which will become a form of environmentally hazardous waste creating environmental and human health-related problems by emitting toxic gas and dust from metal glass, ruthenium, indium, lead, and tellurium\cite{Stewart_Conversation_2019}. 

Given this context, the European Union (EU) and the UK have started working on their regulatory framework, included solar PV waste into a new regulatory directive, and established a legislated recycling policy for solar panels \cite{Majewski_EnergyPolicy_2021} to reduce the continuous growth of PV waste volume by implementing solar module recycling \cite{Chowdhury_ESR_2020}. 
However, legislation alone is insufficient to completely tackle the solar waste problem. For example, EU regulations mandate solar panel recycling, but with current technology, only 75-80\% of materials can be recovered from almost 85\% of collected panels. Partial PV recycling methods mostly recover low-value materials such as glass, making recycling an unprofitable business. The cost-effective recovery of valuable materials in solar panels, such as copper and silver, remains a technical challenge. Japan, Europe, and the USA have initiated research and development initiatives to develop effective solar module recycling techniques. Results from some of these efforts have been reported in \cite{Chowdhury_ESR_2020,Xu_WasteManagement_2018,Majewski_EnergyPolicy_2021}, and \cite{Vargas_JSFI_2021}.

Effective mechanisms to track solar panels' locations, quantity, and EOL once they leave the manufacturing facilities do not exist. The financial and non-financial responsibilities of PV producers, users, and recycling industries are unclear. This present reality of the solar recycling process establishes the need for formal frameworks to track the EOL of each panel, introduce shared responsibilities among different stakeholders, and define how they can make financial contributions to assist the proper recycling process. The proposed CE-driven blockchain-based solution will show how the responsibility of sharing the cost of recycling can be completely (or partially) distributed among the PV users so that they, together with the recycling industries, can contribute to global solar PV waste recycling without significant financial hardship and avoid an unwanted environmental legacy. 

\section{Circular Economy \& Blockchain - Preliminaries}\label{sec:Preliminaries} Before discussing the proposed model in the next section, this section will provide some preliminary understanding of the two key concepts used in this model: circular economy (CE) and blockchain.
\subsection{Circular Economy} A CE can be defined as an industrial system that supplants the EOL concept and encourages the recovery and reuse of materials and products. With a focus on environmental welfare and sustainability~\cite{murray2017circular}, a CE limits natural resource depletion and eliminates the use of toxic waste by introducing new business models incorporating social responsibilities. Through the new business models, it promotes a closed-loop economy by including economic-environmental cooperation \cite{mathews2011progress} and strives to keep resources, goods, and outputs in circulation for as long as possible by keeping their value within the loop \cite{bocken2016product}. However, recycling materials is not always viable as it can be more expensive than discarding them and producing them again from new raw materials. For example, at present, only $10\%$ solar panels are recycled in the USA due to the lack of commercial viability of recycling~\cite{Casey_Recycle_2021}. Since the value of materials from solar panels ceasing operation annually could reach an estimated $2$ billion by the year $2050$, without a proper economic model for recycling will result in a very large amount of unsafe and environmentally unfriendly waste. This will make solar a more environmentally unfavorable piece of the clean-energy puzzle. 

\begin{figure}[t]
\centering
\includegraphics[width=0.5\columnwidth]{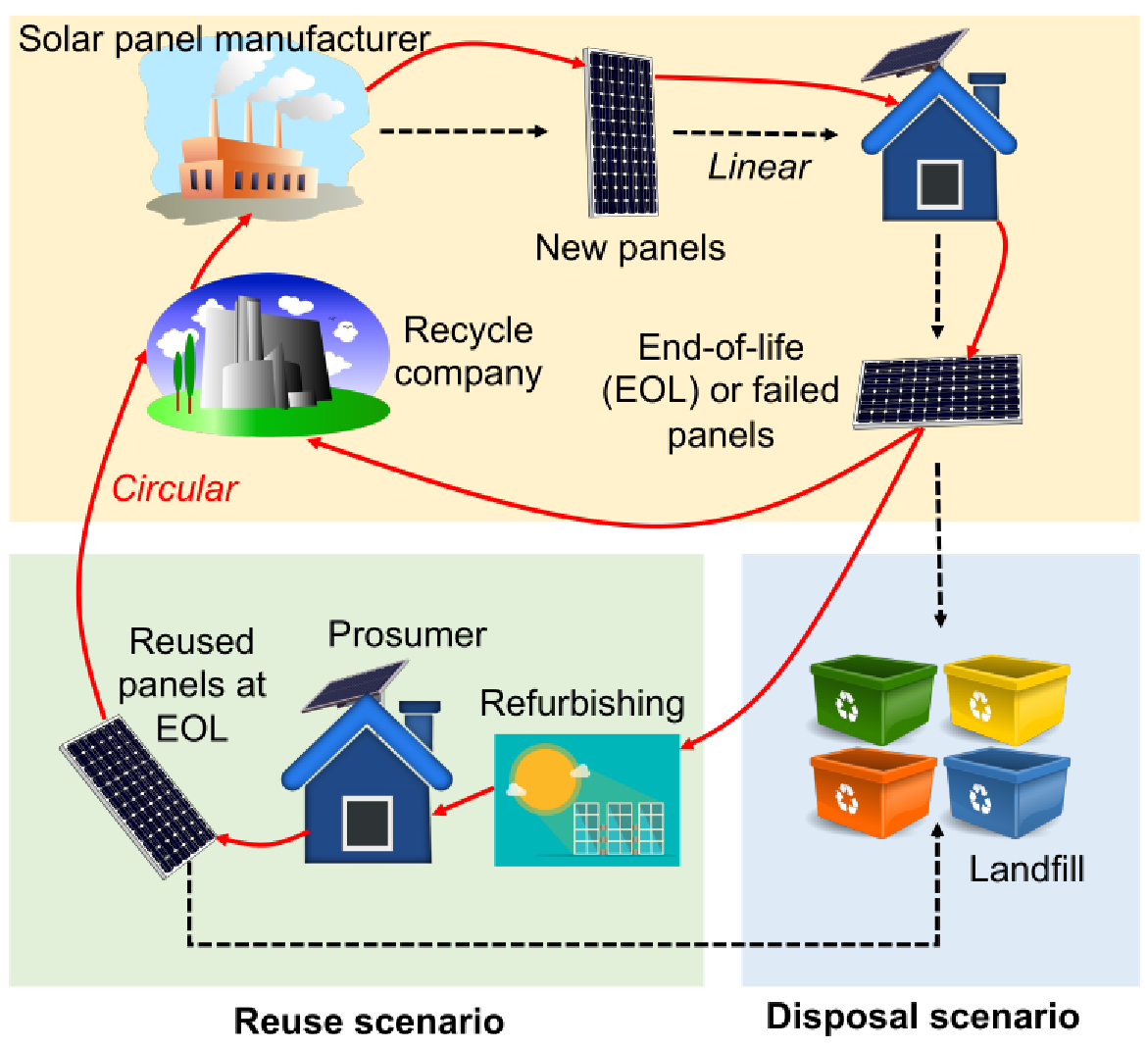}
\caption{The complete life cycle of a solar panel. Black and red arrows, respectively, highlight the linear and circular use. All the components used in this figure are royalty-free and extracted from \url{https://pixabay.com/}}
\label{fig:lincircular}
\end{figure}

A high-level block diagram that shows the complete life cycle of a solar panel is shown in Figure~\ref{fig:lincircular}. This figure is inspired by \cite{Mckinsey_fig}, where the authors describe the life cycle of EV batteries. Red and black arrows highlight the difference between linear and circular use. After leaving the solar panel manufacturing facility, the prosumer purchases a new solar panel. It remains in use until it either reaches its EOL or premature failure. In linear use, the prosumer dumps the solar panel in a landfill. However, in circular use, the solar panel is either sent to a recycling company or a refurbishing company. Another prosumer can use the refurbished panels, and when they reach their EOL they are either disposed of in the landfill or sent back to the recycling company. The recycling company recycles the solar panels and sends the recovered materials to the solar panel manufacturer. This closes the loop and completes the circular life cycle of a solar panel. 

Developing a complete zero-waste solar panel CE requires focus on three essential aspects: circular use, circular recovery, and circular design. Circular use includes the technologies, policies, and business models that enable the sharing of resources and responsibilities among various stakeholders to eliminate waste. Peer-to-peer energy trading, product (solar panels) as a service model, and shared responsibility model are examples of circular use. Tracking the assets (energy production, EOL, cause of failure) and the responsibilities of various stakeholders (users, manufacturers, utilities, and recyclers) for eliminating waste and creating value can be classified as circular recovery. Finally, the circular design includes changes in the solar panel manufacturing process, making panel recycling more manageable and cheaper at their EOL but without compromising their robustness and life span. It also includes the development of low-cost recycling processes for fully recovering all the materials from the solar panels. 

The circular design is an essential component of zero-waste solar panel CE. However, crystalline (first generation) and thin-film (second generation) solar panel manufacturing processes have been standardized, so it is impossible to take this route for the installed and upcoming first and second-generation solar panels. Besides, recycle-friendly changes to the manufacturing processes of these solar panels are also unclear. Therefore, the major focus of the circular design remains on fully recovering materials as much as possible from the solar panels by developing inexpensive recycling processes. This paper, however, does not directly focus on the circular design aspect of CE. Rather, we consider the solar panel circular use and circular recovery aspects for which blockchain provides elegant solutions. The circular design aspect will be indirectly included in our solution through the recycling cost of solar panels, which can be low/high depending on the partial/full recovery and the technological advancements in the recycling process. Table~\ref{tab:CEaspects} summarizes the advantages, challenges, and utility of blockchain for circular use, circular recovery, and circular design. Blockchain is an excellent technology to promote solar panels' circular use and circular recovery. 

\begin{table}[ht]
\caption{This table summarises how blockchain is an excellent technology to promote solar panels' circular use and circular recovery. The circular design is indirectly incorporated into the solution through the inclusion of recycling costs of solar panels.} 
 \centering 
\begin{tabular}{ |p{0.1\linewidth}|p{0.26\linewidth}|p{0.26\linewidth}|p{0.27\linewidth}|}
\hline
 \textbf Aspect & \textbf Description & \textbf Challenges & 
 \textbf Blockchain \\ 
 \hline
 Circular Use & Sharing of resources (energy, assets) and responsibilities (costs, revenues) among various stakeholders to eliminate waste & 
 Enabling participation, lowering barriers, enhancing trust & Good match as blockchain is inherently distributed and allows sharing of information more seamlessly.  \\
  \hline
 Circular Recovery & Tracking of assets (solar panels, energy production, EOL, cause of failure) and responsibilities of stakeholders & 
 Distributed nature of assets and resources, variations in usage, quality, lifetime & Good match because blockchain allows all the actors (with appropriate access privilege) to track the information/asset. \\
  \hline
  Circular Design & Changes in the solar panel manufacturing process to make recycling valuable and cheap & 
 Mature solar panel manufacturing processes, enhancing panel life while making them recycle friendly & Poor match as blockchain cannot contribute directly to recycling but bring together multiple competing stakeholders to enable the circular economy. \\
  \hline
\end{tabular}
\label{tab:CEaspects} 
\end{table}

Circular use can be achieved through responsibility sharing. For example, producers and consumers of green energy could take responsibility to help the recycling industry recycle their solar assets. Recent studies found that people if trained and educated appropriately, are willing to pay more for bearing the cost of using renewable energy resources~\cite{Mengelkamp_EP_2019,Hackbarth_EP_2021}. Few other researchers have discussed collective recycling responsibility in supply chains. For example, Nie et. al \cite{nie2013collective} have proposed a recycling responsibility-sharing model for the fashion industry. Khedlekar and Singh have presented a three-layer supply chain policy under the sharing recycling responsibility model in \cite{khedlekar2019three} for the manufacturing sector. Both papers have examined the effect of responsibility sharing a percentage of a manufacturer, a retailer, and a collector. 

Therefore, appropriate shifting of financial responsibilities for recycling to the green energy producers and consumers could be a feasible solution to the global solar waste recycling dilemma. However, the burden on them should not be prohibitive. At the same time, tracking the assets and responsibilities for circular recovery are also crucial. This way, the CE can create cooperative and transformative business strategies to ensure sustainability through blockchain-based digitalization, as shown later in this article.

\subsection{Blockchain} 
\begin{figure}[t]
\centering
\includegraphics[width=0.99\columnwidth]{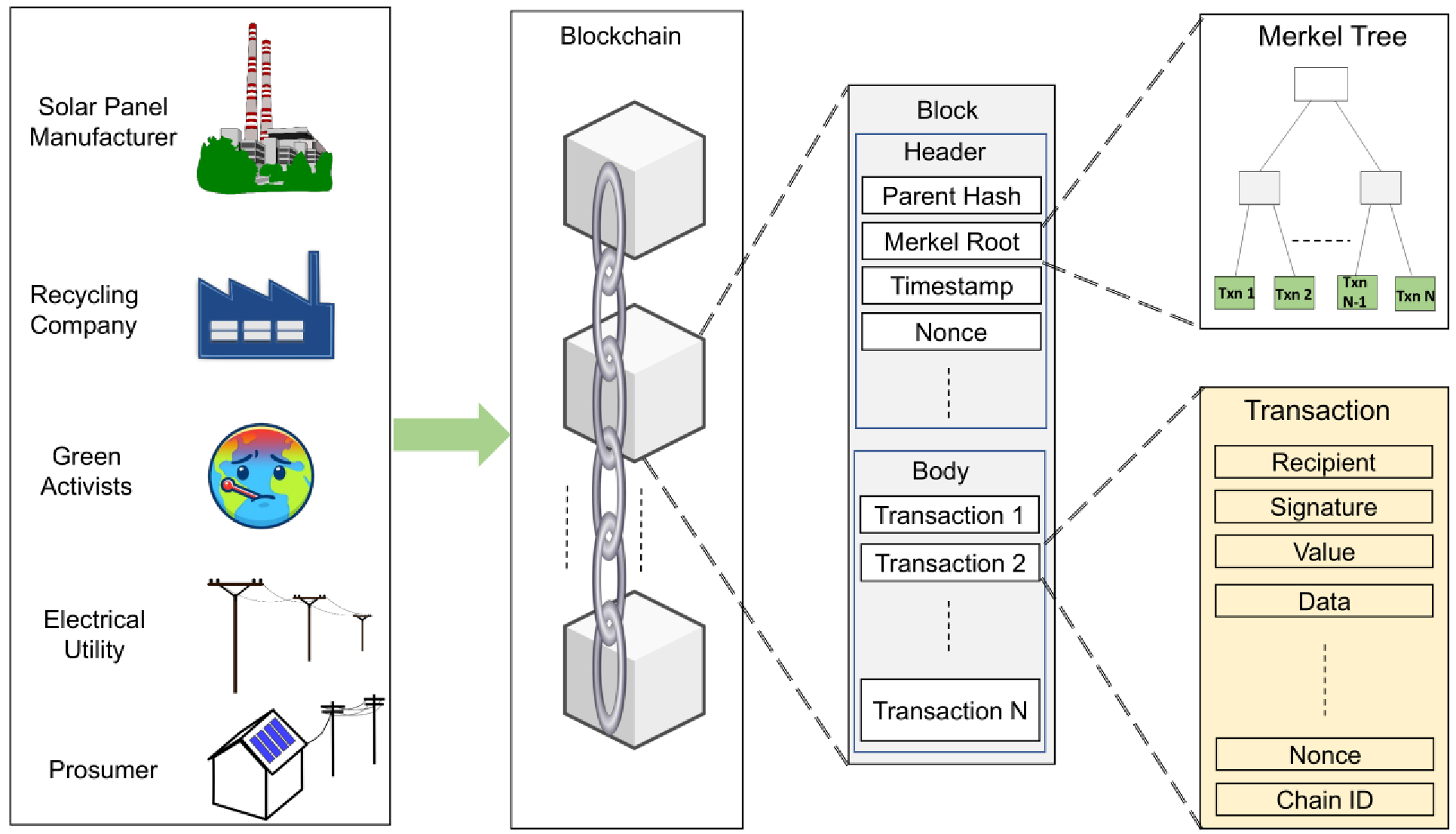}
\caption{A conceptual diagram showing the actors in our CE (solar panel manufacturer, recycling company, green activists, utility company, and prosumers) interacting with a Blockchain. Further, the contents of a block, transactions, and the process of creating a Merkel tree are also illustrated. All the components used in this figure are royalty-free and extracted from \url{https://pixabay.com/}}
\label{fig:blockchain}
\end{figure}

A blockchain is a decentralized ledger or a continually updated list of transactions that record agreements, contracts, and sales \cite{Naveed_IEM_Dec_2019}. The fundamental component of a blockchain is a block, which contains a header and body - some content under the header. The header always contains the hash of a previous block header (parent block) to create a cryptographic link and a tamper-proof chain. Several other fields with unique purposes such as a timestamp~\cite{Massimo_CSE_Sept_2017}, a Merkle tree root~\cite{DonghyeokLee_MTA_2021}, Proof of Work (PoW)~\cite{ferdous2021survey}, and nonce~\cite{LeilaIsmail_ICBCT_2019} may also be included in the block header. The block body contains transactions where useful data/information is stored. The transactions are created and signed by blockchain users. A conceptual blockchain diagram is shown in Figure~\ref{fig:blockchain}.

Public key cryptography is used in blockchain, where every user/account is assigned a keypair (private and public key)~\cite{Michael_AIR_June_2016}. The private key is known only to the user, while the public key is known to everyone in the system. A user requires its private key to create and sign new transactions and provide proof of asset ownership. Other nodes (blockchain users) in the network verify the transactions and claims of a user with the help of its public key. The nodes in a blockchain network only propagate transactions that satisfy the specified rules (valid transactions) and drop all invalid transactions. At regular intervals, multiple transactions are picked up by a few special nodes called miners or validators and packaged into a new block. The miner or validator creates a block header that satisfies the blockchain's consensus algorithm criterion. For example, in the PoW consensus algorithm, the miner repeatedly tries various values of the nonce (field in the header) until the block header hash value is less than a specific target value. The new block with a valid header (and valid transactions as the body) is disseminated in the network. Other miner nodes verify the transactions and the header value in the proposed block. If this block satisfies all the rules, all the miner nodes append the new block by linking it as the head of the existing blockchain. The nodes generally follow the longest chain rule to resolve all the natural forks. Once a block becomes part of the blockchain and has received a sufficient number of confirmations (child blocks), it becomes extremely difficult to change its contents because, due to chaining, the hashes of all the child blocks will also change and would have to be computed again requiring a lot of effort (in terms of resources) making it practically infeasible for the attacker.   

A public blockchain is called a permissionless blockchain, where anyone connected to the Internet can join the blockchain network and become a part of it with the ability to create and audit transactions. On the other hand, a private blockchain is called a permissioned blockchain, in which a user requires permission from the system administrator to join the blockchain network. A private blockchain is owned and maintained by one organization, and robust access control is maintained to restrict access to the network to only particular members. A consortium blockchain combines private and public blockchain features, where more than one party/organization is in charge. A consortium of organizations acts as a gatekeeper and provides or revokes access to nodes for reading, writing, and auditing the blockchain. Every organization in the network has access to the data based on the agreed access control rules set by the consortium. Since no single authority governs the control, it maintains a decentralized nature.  Consortium blockchain also supports higher scalability. For example, the Hyperledger Fabric shows a good transaction throughput to 30000 transactions  \cite{gorenflo2020fastfabric}.
\subsection{Blockchain-based Circular Economy} Recently, researchers have explored how blockchain can be used in the context of CE \cite{kouhizadeh2020blockchain, upadhyay2021blockchain, yildizbasi2021blockchain, narayan2020tokenizing}. The CE can benefit from simultaneous cooperation and competition by adopting a decentralized principle of value creation, and circulation instead of value creation and value appropriation \cite{narayan2020tokenizing}. Thus,  blockchains, with their immutability of the record and data flow transparency, can help achieve the CE's goal. The CE's goal to reduce or eliminate waste can be accomplished by converting EOL products/services into restorative value by enhancing longevity or deciphering new forms of consumption~\cite{sousa2017lifecycle}, allowing them to circulate sustainably. Blockchain can help devise a new business model with multiple stakeholders in the circular economy to ensure the cyclical life cycle in product and service design. Blockchain can also contribute by providing the foundation technology for connecting complicated networks and databases with built-in designs and settings to update all linked databases concurrently, irreversibly, and automate when necessary. In addition, the smart contract \cite{zheng2020overview} and tokenization \cite{narayan2020tokenizing} can automate the process and control or document legally relevant events and actions according to the terms of a contract or an agreement by multiple parties. 

The following section presents the circular economy-driven blockchain-based solar recycling solutions. For that purpose, here we introduce smart contracts, full nodes, digital wallet applications, and escrow accounts that are used in the solution description. Smart contracts~\cite{Rafael_IEEECommSM_Sept_2018} are computer programs that are stored on a blockchain and executed when some prerequisites are satisfied. Smart contracts are used to automate the process of executing an agreement so that the involved parties can agree on outcomes without a central authority. A full node~\cite{Wenbo_Access_Jan_2019} (often called miner node) maintains the integrity of the blockchain network by storing a complete copy of the blockchain. This node can also verify and add new blocks to the blockchain by following some consensus mechanism (e.g., PoW or PoS). There could be other participants in the blockchain network who are not full nodes but can access the blockchain-based application. Digital wallet application (DApp)~\cite{Mohamed_IOT_Apr_2019} is a software tool that allows the user to interact with the system. It also stores, sends, and receives digital currencies/tokens. This software creates transactions on behalf of the user and posts them on the blockchain. Finally, escrow account~\cite{Rafael_IEEECommSM_Sept_2018} is a type of account in blockchain that can hold tokens and only release the tokens to the appropriate party if a specific condition - written as a smart contract - is satisfied. The parties involved in the transaction need to ensure that the agreed product/service is delivered and payment is made. 

Different researchers have explored and analyzed the use of blockchain for circular economy. Bockel et al. have presented a details analysis of the research and practice gap in using blockchain for the circular economy in \cite{bockel2021blockchain}. They have highlighted three key issues: 1) Research lacks a well-defined terminology outlining the various types of blockchains, their technical attributes, and associated advantages, 2) Establishing trust and verification poses both significant potential benefits and challenges, and 3) A thorough exploration of the potential advantages and challenges posed by blockchain technologies in the context of the circular economy, intertwined with sustainable development, is imperative. Corsini et al. have explored how blockchain can support the development of circular smart cities and measurement tools for providing information to stakeholders and assisting in policy creation for circular economy \cite{corsini2023fostering}. Shojaei et al. \cite{shojaei2021enabling} have presented and tested through a synthetic case study to provide a proof of concept as to the feasibility of blockchain as an enabler of a CE in the built environment. 

\section{Circular Economy-driven Blockchain-based Solar Recycling}\label{sec:ProposedModel} In a solar-based renewable energy system, multiple parties are involved including regulator and green activists, solar panel manufacturers and installers, utility companies, solar panel recyclers and refurbishers, and solar energy producers. At present, solar energy producers buy solar panels from the manufacturer with an up-front investment, enjoy the clean energy for domestic use for free \cite{Isak_ERSS_June_2020}, earn revenue by selling surplus electricity either to the grid or to neighbors \cite{Tushar_TSG_July_2020}, and can participate in more welfare activities by donating electricity for charitable purposes \cite{Mahdi_AE_Jan_2022}. And, solar recycler takes responsibility for the panels once they reach their EOL and do not end up in the landfill. 

In the traditional model~\cite{sanderink2020institutional}, stakeholders do not interact when promoting the use of solar energy. There is no trusted centralized entity to coordinate activities and enforce the responsibilities of various parties to create a CE of solar panels. Additionally, there is no centralized tracking mechanism for solar panels at different stages of their life cycle. For example, solar panels can be manufactured in one region, used in another region, and recycled in another different place. On top of this, solar panel recycling is not profitable for recyclers and would require cost-sharing solutions. The inherently decentralized nature of this problem calls for a decentralized solution that should allow non-trusting parties to create the CE of solar panels. 

This paper presents blockchain-based decentralized solutions to support solar panels' CE because blockchain can bridge the trust deficit between the energy producer and consumer. Secondly, smart contracts in blockchain can transparently automate the responsibility-sharing function as all the parties in the system have access to all data related to them. The proposed solutions vary in their complexity and utility of blockchain features. We assume that all the parties in the system, i.e., the prosumer, manufacturer, and the recycling company, agree on the recycling cost per module and solar panel lifetime. These values are recorded on the blockchain. The proposed model for sharing the responsibility for recycling uses a blockchain-based platform for real-time tracking of the use of solar energy by different stakeholders and transacting the reasonable cost of recycling across different accounts over the lifetime of the solar panels. 

The proposed model assumes a consortium blockchain, where solar panel manufacturers, recycling factories, utilities, and governments/regulators are full nodes that store the full blockchain and participate in the block creation. Due to capital investment constraints, energy producers and consumers are considered light nodes that can use the system and create transactions. Further, we assume that they are connected to the network via DApp wallet, which allows them to access their relevant information. In the proposed model, the roles of different stakeholders within a solar system can be summarised as follows:
\subsubsection{Regulator/Green Activists} The regulatory body and the green activists (private NGOs) monitor the compliance issue and ensure that all participating stakeholders abide by the regulations. They do not contribute and share the cost of recycling but maintain a full blockchain node, which allows them to audit the transactions.
\subsubsection{Manufacturer} The manufacturer is responsible for producing the solar panels and supplying them to the consumers. During the warranty period, the manufacturer is responsible for replacing faulty panels and reusing the recycled panels for future manufacturing.
\subsubsection{Utility company} The utility company is responsible for providing smart meters and keeping track of the energy production, consumption, generation, and collection of bills. 
\subsubsection{Recycling and Refurbishing Companies} These are responsible for recycling or refurbishing the panels from the solar plants and sending the recycled panels back to the manufacturer for reuse. The cost of the recycling process is carried out by the money accumulated in the escrow account and funding from the Government to complement the gap between the total cost and the balance in the escrow account.
\subsubsection{Consumer/prosumer} Solar panels of a prosumer generate green energy that can either be self-consumed or sold to neighbors (peer-to-peer) and the grid. The prosumer either earns revenue by selling energy or benefits from self-consumption through a reduction in energy bills. We assume that after a certain number of solar energy units are generated and recorded by the smart meter, the prosumer pays an additional contribution towards solar panel recycling that is accrued in an escrow account. Given the long lifespan of solar panels (25 to 30 years) and recycling costs in the range of \$15 to \$45 per module (0.0428\$/W - 0.125\$/W)~\cite{USDep_2021}, these contributions are not expected to create much burden for the prosumer (a few dollars per year). When the solar panel reaches its EOL, the total amount accumulated in the escrow accounts is split into two. One part will go to the recycler to meet the cost of the recycling and send the recycled panels to the manufacturer. The rest will be credited to the prosumer to award the producer for transferring the expired solar panels to a recycling facility rather than to the landfill. Privacy is a concern for such a system. The privacy of the prosumers can be protected by using the private channels in a blockchain system.  In the consortium blockchain, the private channel is used to keep the communication private between concerned/related parties, e.g., energy producers and consumers. The manufacturers and recycling companies can still see the gross energy generation/consumption by the prosumers to maintain the transparency of the system. In the permissioned blockchain network like Hyperledger Fabric, separate channels for interactions between different organizations can be created. Access to the channel (as well as to the overall network) is controlled via the certificate authorities and cryptography-based authorization. Channels allow organizations to use the network while maintaining separations between multiple blockchains. Only the members of the channel (including the stakeholders and Observer) can see the specifics of transactions performed on that channel, whereas other members of the network cannot see the transactions on that channel \cite{yang2020public}.

To elaborate on how blockchain can be used in different ways to share the cost of solar panel recycling among different stakeholders, we propose three solutions. These solutions differ from one another depending on how blockchain is being utilized as a platform in the solution and terms of their advantages and limitations. We note that one limitation of using blockchain is that it consumes a substantial amount of energy. However, we expect that as the concept of shared responsibility of solar panel recycling will become more common, more innovations will be made through research and development initiatives to address this issue.

\subsection{Solution 1 - Blockchain for Information and Data Storage Only} 
\begin{figure}[t]
\centering
\includegraphics[width=0.99\columnwidth]{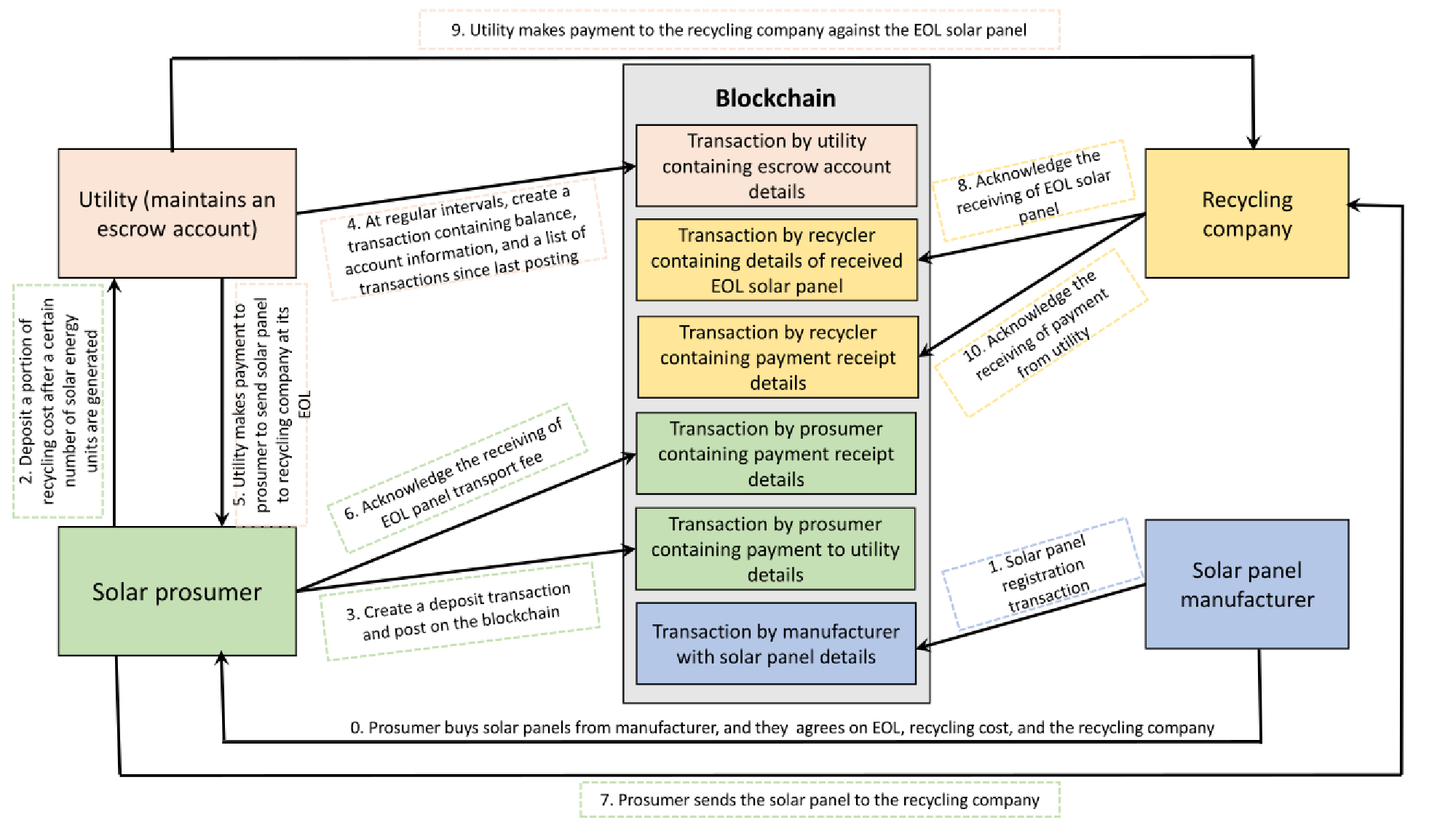}
\caption{Solution 1: This figure demonstrates how different entities within a solar system are connected to a blockchain platform and contribute to recycling the solar panel through the circular economy concept.}
\label{fig:sol1}
\end{figure}

In the first solution, we use blockchain for information and data storage, while the financial settlements are performed outside the blockchain in fiat currencies. The actions of various parties in terms of their financial obligations are stored on the blockchain as transactions. The inherent properties of blockchain, such as data immutability, transparency, and non-repudiation, help build trust among various entities. A high-level conceptual diagram is shown in Figure~\ref{fig:sol1}. 

We propose a consortium blockchain controlled by utility companies, solar panel manufacturers, and recycling companies and these entities are responsible for maintaining the blockchain and validating the transactions (we call them validators). The prosumers are also part of this blockchain, can create transactions, and take part in validating some transactions. However, prosumers cannot create new blocks. We assume that every node (prosumers and validators) has a unique address, and a public/private key pair is assigned to it for creating and signing transactions. We assume that any node in the network, including the prosumers, regulators, and green activists who may also be given access to the blockchain, can read and audit the information in various blocks. 

When a prosumer buys solar panels from a manufacturer, they mutually agree on the EOL, recycling cost, and the recycling company. Additionally, they also agree on the warranty claims. The manufacturer creates a new transaction that contains this information and permanently stores it in the blockchain. After a certain number of energy units are produced by the solar panel, the prosumer pays a portion of the recycling cost to the utility company, which deposits it in a bank account. The utility company is at the forefront of this model. It collects the recycling cost and the energy bills (if prosumers need to buy energy from the utility company) from the prosumers. The utility company also provides the smart meter to the prosumer and thus keeps track of the energy units generated by solar panels. The prosumer creates a transaction for this payment and records it on the blockchain. The prosumer also creates a new transaction when the solar panel reaches its EOL and receives payment from the utility company for transporting the solar panels to the recycling facility.    

To track the balance of the recycling bank account, the utility company creates transactions and posts them on the blockchain at regular intervals. This transaction contains the bank account number, current balance, and active customers contributing to this recycling account. These transactions, along with those made by the prosumers, allow anyone on the blockchain to verify the financial contributions, state of recycling funds, and a check on the utility company that keeps custody of the physical dollars. The utility company only creates these account balance transactions.

 When the recycling company receives solar panels from the prosumer, it creates a transaction on the blockchain and indicates its readiness for recycling. The utility company pays the recycling company from the bank account it maintains. The recycling company creates a new transaction after receiving the payment. This concludes the normal life cycle of a solar panel where it reaches its EOL. In this case, the recycling cost is completely recovered over the solar panel's entire lifespan (25 years). 

Some solar panels can also fail prematurely before reaching their normal EOL. In this case, the recycling cost is not fully recovered through solar panel usage. We consider two failure cases: 1) Solar panel fails during the warranty period before its expected EOL, and 2) Solar panel fails after the warranty period but before its expected EOL. It is worth mentioning that without blockchain administering such cases and attributing responsibility can become challenging due to a lack of trust in the utility company or a lack of transparency without an immutable audit trail. 

\begin{figure}[t]
\centering
\includegraphics[width=0.99\columnwidth]{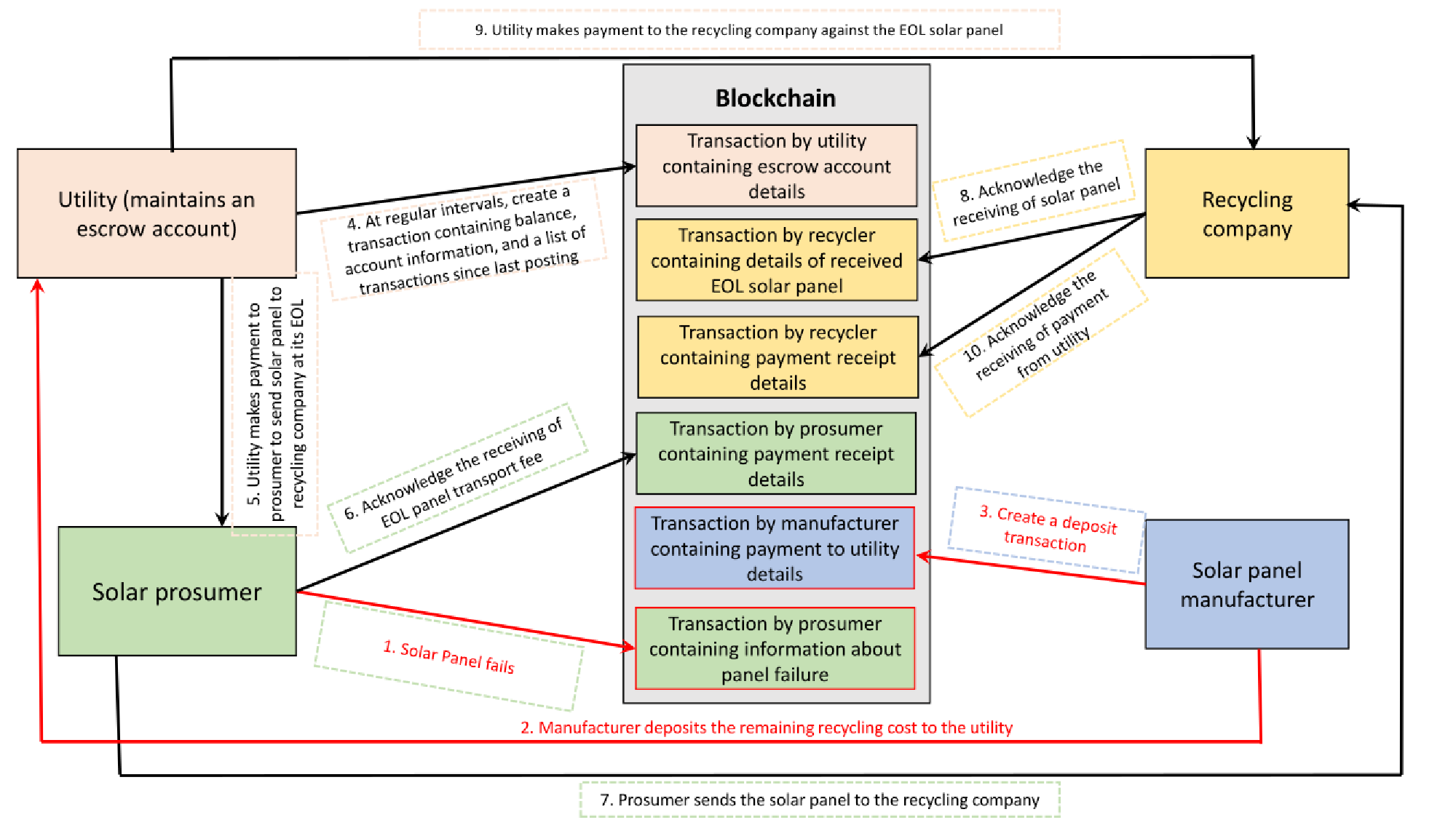}
\caption{This figure demonstrates how the proposed solution can be used in the case where solar panels fail within the warranty period and before reaching their EOL.}
\label{fig:Warr1Sol1}
\end{figure}

If the solar panel fails before the warranty period, the manufacturer is responsible for replacing the solar panels and covering the remaining recycling cost. In this case, the prosumer creates a transaction containing information about the cause of failure, the remaining recycling cost, the warranty claim of the prosumer, and the panel and manufacturer details. The manufacturer deposits the remaining recycling and transportation costs in the bank account and creates a transaction for this payment on the blockchain. Once the utility company receives the complete payment, the remaining steps are the same as we discussed in the normal EOL case. The utility company pays the prosumer the transportation cost, the prosumer records this transaction on the blockchain, then ships the panels to the recycling company and records another transaction on the blockchain. The recycling company receives the solar panels and creates a transaction acknowledging solar panel receipt, the utility company makes the payment to the recycling company, and the recycling company acknowledges by creating a transaction for this payment. The steps different from the normal EOL operations are highlighted in (red color) Figure~\ref{fig:Warr1Sol1}. The use of blockchain creates trust among stakeholders because the data is transparent and immutably stored on the ledger.

\begin{figure}[t]
\centering
\includegraphics[width=0.99\columnwidth]{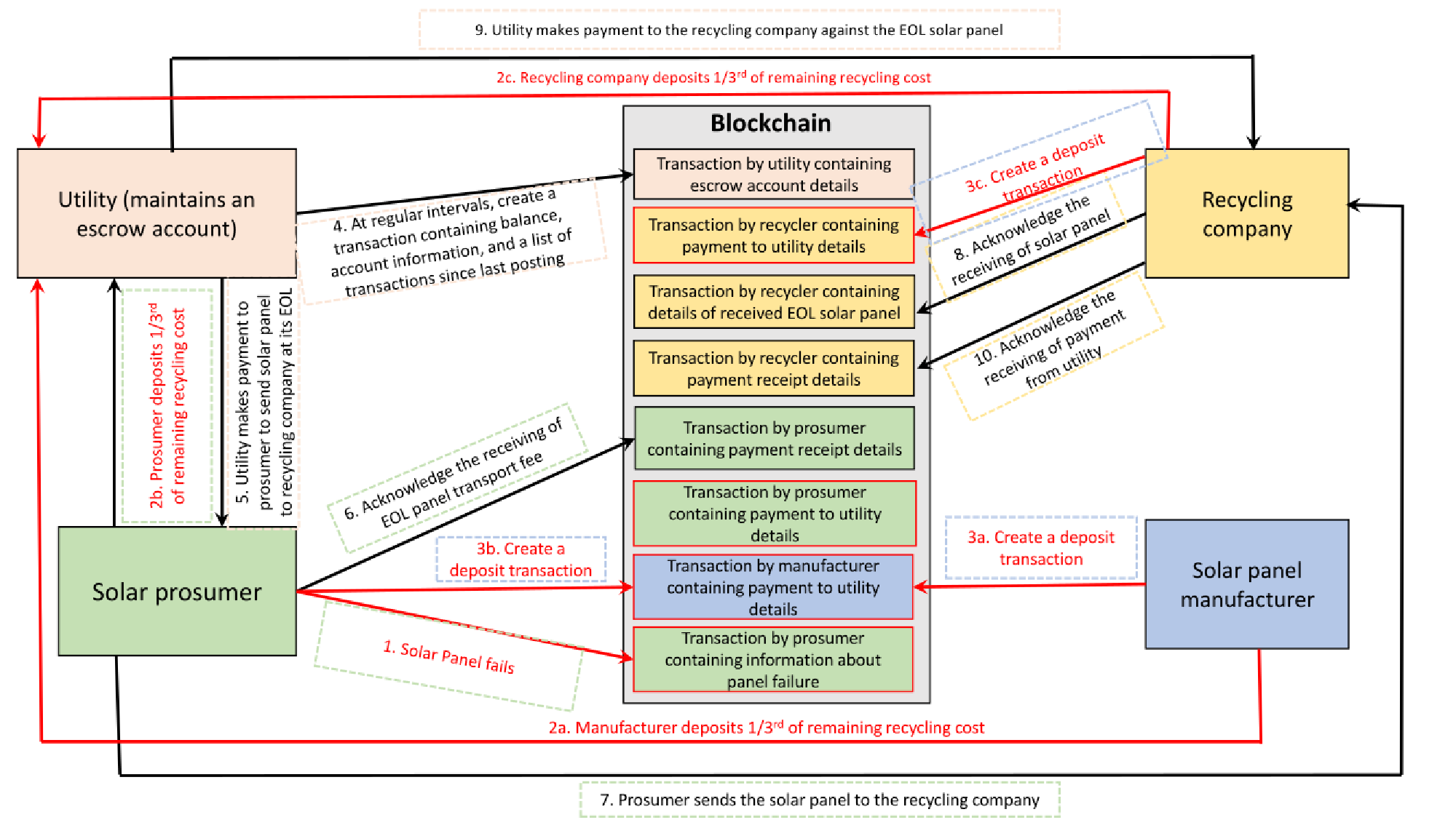}
\caption{This figure demonstrates how the proposed solution can be used in the case where solar panels fail after the warranty period but before reaching their EOL.}
\label{fig:Warr2Sol1}
\end{figure}

In the other case where the solar panel fails after the warranty period but before its EOL, prosumers may be interested in discarding the solar panel without any further payments since there are no benefits of clean energy from these panels. In this case, we propose a solution where the responsibility is equally shared among the manufacturer, the prosumer, and the recycling company. The prosumer creates a transaction containing information about the cause of failure, the remaining recycling cost, the warranty claim (zero in this case), and the panel and manufacturer details. The prosumer, the manufacturer, and the recycling company deposit one-third of the remaining recycling cost each to the bank account and create their respective payment transactions. This one-third of the payment used in this case is just a numerical example to illustrate the idea. This may further be optimized according to the system requirement. Once the utility company receives all the costs, the remaining steps are the same as we discussed in the normal EOL case. The prosumer has a strong incentive to pay the reduced liability because its identity is known in the system, and non-payment can create difficulties for its future solar panel purchase. The steps different from the normal EOL operations are highlighted in (red color) Figure~\ref{fig:Warr2Sol1}. Again, blockchain is the key enabling technology that facilitates handling such cases with relative ease.   

Our model creates very little financial hardship for the prosumer. Assuming a high module recycling cost of 0.125\$/W (\$45 per 350 W module) and EOL of 25 years, the prosumer pays 0.005\$/W/year, equivalent to 0.0004167\$/W/month. Suppose a prosumer has 10 kW solar panels installed on the rooftop; then the monthly contribution becomes \$4.167. On the other hand, if the recycling cost is included in the solar panel cost at the installation time (initial purchase), then the prosumer would be required to pay an additional $\$1285$, which will create a significant financial hardship. Moreover, in our approach, linking the payments with the energy produced by the solar panels further limits the hardship for the prosumer because when the panels produce more energy, the prosumer enjoys more benefits either in terms of revenue from selling additional energy units or in the form of reduced energy bills, and therefore would pay more towards the recycling cost. Thus, depending on the seasonal variations in weather at a given geographical location, the contribution towards solar panel recycling would also be low in periods of low energy production.

Using blockchain to create this CE for solar panel recycling has several benefits. Blockchain enables tracking assets, EOL, funds, and responsibilities. At every step, transactions are created by various players in the CE, which are very helpful in case of disputes. Blockchain also allows regulators and green activities to audit the activities of various actors in the system. The detection of malicious activities or malicious actors is straightforward on the blockchain because the transactions are signed by respective entities and can be easily analyzed. In addition, as we are using consortium blockchain, it will keep us safe from other common types of blockchain attack/vulnerability such as 51\% attack and Sybil attack.

In Solution 1, we use blockchain to record data. Implementing this solution is easy, but it does not leverage the full potential of blockchain. Moreover, this solution lacks automation, and the utility company is in physical possession of the recycling fee and handing out payments to the prosumer and the recycling company at the end of the solar panel EOL. In the next subsection, we present another solution that relies on smart contracts and stablecoins and is more secure but relatively more complex than Solution 1.   

\begin{figure}[t]
\centering
\includegraphics[width=0.99\columnwidth]{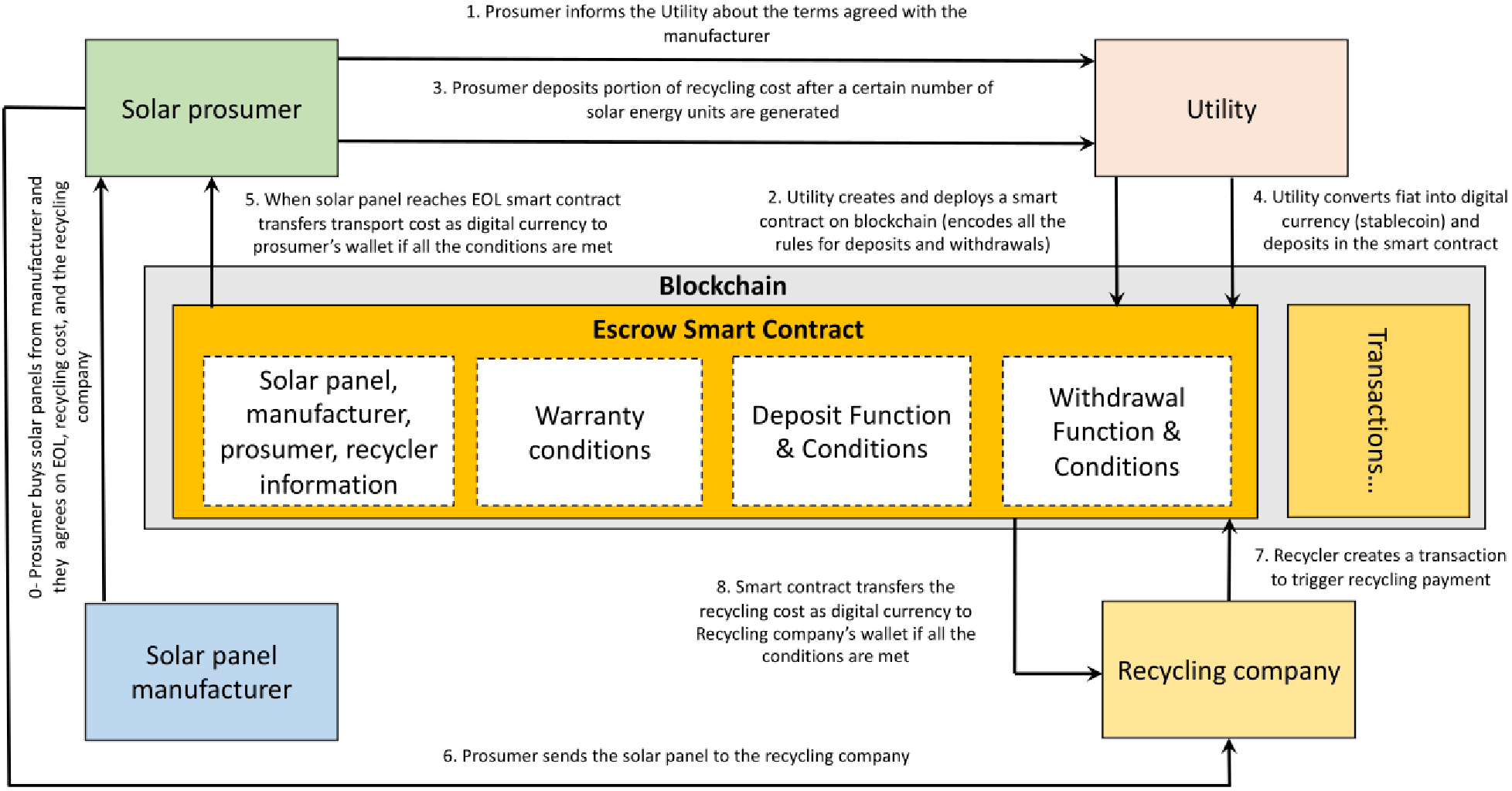}
\caption{Solution 2: Automation is achieved through Smart Contracts, and Recycling Fee is stored on Blockchain in the form of Digital Currency held by the Smart Contract.}
\label{fig:Solution2}
\end{figure}

\subsection{Solution 2 - Smart Contract and Digital Currency enabled Solution} Blockchain is more than just information storage in this solution. This solution requires smart contracts and provides more automation. Here, the prosumer also deposits the recycling fee and electricity bill (if needed) to the utility company. Hence, this part is as same as Solution 1. However, there are no bank accounts, and the recycling fee is stored on the blockchain as digital currency (stablecoins~\cite{StableCoin_FRL_2021}). The utility company deploys a smart escrow contract on the blockchain and encodes all the information about solar panels, manufacturers, recyclers, EOL, warranty claims, deposit rules, and withdrawal rules. Most of the transactions discussed in the previous solution are also retained because these transactions trigger certain smart contract functionalities. 

The prosumer pays the recycling recycling cost to the utility company over the lifetime of the solar panels, and the utility company converts it into digital currency and deposits it into an escrow smart contract. When the solar panel reaches its EOL, and if all the required conditions are met as encoded in the smart contract, the prosumer's wallet is credited by the smart contract with the digital currency equivalent to the transportation cost of the solar panels. The prosumer then ships the solar panels to the recycling company. Upon receiving the solar panels, it generates a transaction to trigger the recycling fee withdrawal from the smart contract. Once again, if all the precoded conditions are satisfied, the wallet of the recycling company is credited with an equivalent amount of digital currency.

Regarding liabilities, cases where the solar panels fail before reaching their EOL will be dealt with in the same way as in Solution 1. The manufacturer, recycling company, and the prosumer will pay the utility company in fiat currencies, which will convert into digital currencies and be deposited in the escrow smart contract. In this solution, digital currency is a one-to-one mapping means, \$1 is equivalent to 1 unit of cryptocurrency (similar to a stable coin in the cryptocurrency world). Although the stability of the price of the stablecoin may vary due to inflation or other market dynamics, we can peg our coin against solar panel recycling costs to reduce the volatility. Furthermore, we can peg them against government-issued digital currency (e.g., China \cite{xu2022developments} ) in the future. Pegging against this kind of government-issued digital currency (often called Central Bank Digital Currency (CBDC)) will stabilize the price of the stablecoin and reduce the volatility of our stablecoin. Once all the parties settle the liabilities, the smart contract will make payments to the prosumer and the recycling companies in digital currencies as we already described for the normal EOL case.  

This solution provides better security and automation because the recycling fee is converted into digital currency, and the actual value and information are stored on the blockchain. We can enhance the security of this solution by using multiple signatures to trigger withdrawals. For example, the smart contract may require the signatures of the prosumer, the recycling company, the manufacturer, and the utility company to transfer the digital currency. All the signatures would indicate the actors' consent before the smart contract makes the payments. A high-level conceptual diagram of Solution 2 is given in Figure~\ref{fig:Solution2}.  

\subsection{Solution 3 - Creation of Digital Recycling Economy Using DeFi} 
This solution uses Decentralized Finance concepts and is more advanced than the previous ones. Decentralized Finance, often abbreviated as DeFi, refers to a category of financial services and applications built on blockchain technology by enabling peer-to-peer transactions without the need for traditional financial intermediaries such as banks or brokerages~\cite{zetzsche2020decentralized}. DeFi platforms operate on decentralized networks like Ethereum, utilizing smart contracts to automate and execute transactions without the need for intermediaries. DeFi platforms leverage smart contracts, which are self-executing contracts with the terms of the agreement directly written into code. This enables the automation of various financial functions such as lending, borrowing, trading, and asset management.

An entirely new coin called a `recycling coin' (RC-coin) is created to incentivize solar panel recycling. Instead of prosumers paying recycling costs in the energy bills, a certain number of so-called RC-coins are produced whenever the solar panels of a prosumer generate a certain number of energy units. The utility company can verify this generation of green energy units by the solar panels of a prosumer in the same way as in solutions 1 and 2 (the smart meter sends this information to the utility company for billing purposes). The prosumer creates corresponding transactions (Coin-minting transactions) and deposits the coins in an escrow contract deployed on the blockchain. Our new coin derives its value from the green energy produced by solar panels. Like any cryptocurrency, RC-coins are also traded on crypto exchanges, and any user can buy or sell these coins. This native token (RC-coin) can incentivize users to participate in platform activities, such as providing liquidity, staking, or participating in yield farming. Users receive rewards in the form of additional tokens or a share of transaction fees for contributing to the platform's liquidity and security.

We assume $W$ RC-coins are minted for every $E$ energy unit produced by the solar panels. The number of generated coins depends on several factors, including the solar panel recycling fee agreed between the parties, coins in circulation, and the price stabilization mechanism implemented in the DeFi protocol. The prosumer deposits a certain portion $\gamma W$, where $0 < \gamma \leq 1$, of these coins is deposited in an escrow contract deployed on the blockchain. This smart contract is reserved for payments to the recycling companies (same functionalities as the recycling fee smart contract in solution 2). The remainder goes into a second smart contract which we call the `Reserve currency smart contract' (RCSC). This smart contract acts like a reserve bank to control the liquidity and price of RC-coins. RCSC is like a reserve bank that manages digital currency reserves. The inspiration for RCSC comes from a central bank in traditional financing, which monitors and regulates the currency. Here, RCSC is a smart contract and the code itself is in control. The regulation policies are encoded and enforced by RCSC. RCSC will enforce the currency regulation policies. However, unlike a central bank in traditional finance, the policies are enforced through the programmed instructions deployed on the blockchain. The solution will remain decentralized because no party on its own in the recycling circular economy will be able to influence the smart contract.

Recycling fee payments are made from escrow smart contracts in RC-coins. However, the coin price fluctuates according to the market dynamics, and it can be controlled to some extent by the liquidity and price control mechanisms embedded in the DeFi protocol. The smart contract will convert the agreed-upon recycling fee into an equivalent number of RC-coins and transfer them to the digital wallet of the recycling company when the solar panels reach their EOL. Similarly, the smart contract will convert the transportation cost into an equivalent number of RC-coins and credit the digital wallet of the prosumer. In case of pre-mature failures before the EOL, the liabilities will be settled in RC-coins. For example, suppose the solar panels fail before the warranty period. In that case, the manufacturer will purchase RC-coins equivalent to the remaining recycling fee from the market and deposit them into an escrow smart contract. Similarly, suppose the solar panels fail after the warranty period. In that case, the prosumer, the manufacturer, and the recycling company will purchase RC-coins from the market equivalent to their portion of the liability and deposit them in the escrow smart contract. 

The number of installed solar panels is increasing yearly along with the number of energy units produced by them. According to some estimates, the global installed solar PV capacity has already reached 1TW. In 2021, the total energy produced by all these solar panels was around 1000~TWh, equivalent to 1 trillion energy units (1T kWh). In our model, RC-coins are produced after a certain number of energy units are generated by the solar panels of a prosumer. Therefore, the number of RC-coins will continuously increase, creating a greater risk of coin oversupply that would negatively impact its value. On the other hand, it is also essential to maintain sufficient liquidity to ensure the coin price does not become excessively high and lose its operational value for micro-payments. We can adopt several mechanisms to control the coin supply and embed them in our DeFi protocol. We discuss two mechanisms without giving excessive details.

The idea is to produce the same number of new RC-coins every year despite a consistent increase in the overall solar energy produced by the solar panels. This can be done in two approaches. In the first approach, we fix the total number of energy units, after which new RC-coins are minted but decrease their quantity. For example, let us assume that in year N, the DeFi protocol awarded 100 RC-coins for every 1000 energy units produced by the solar panels. Suppose the overall energy units produced by all the solar panels in the system are expected to increase by 25\% or a factor of 1.25 in year N+1. Therefore, the protocol will now reduce the award to 100/1.25 or 80 RC-coins per 1000 energy units—this way, the same number of RC-coins will be produced yearly. In the second approach, we can effectively maintain a constant coin generation rate by increasing the number of energy units while keeping the number of RC-coins constant. For example, in year N+1, the protocol can award 100 RC-coins per 100x1.25 or 125 energy units. Both approaches will achieve the same result.  

Like the central bank, the RCSC, in our solution, can employ various strategies, which can be encoded in the form of complex procedures to regulate the coin price and its supply in the market. We can allow the protocol to burn some RC-coins it is holding and take them permanently out of circulation depending on the market capitalization and the circulating supply. Like a central bank, RCSC can also buy and sell coins in the market. For example, there may also be users/entities wanting to support the CE of solar panels without owning them. Such users may purchase RC-coins from the market and donate them to RCSC. We can also allow RC-coin transfer from RCSC to the escrow smart contract under certain conditions. Despite these measures, the price of the coin will fluctuate depending on market dynamics, making this solution challenging and interesting. In this paper, we are only presenting an overall concept of this solution without going into many technical details because the values of several parameters and protocols have to be carefully thought out to properly roll out such a solution and ensure the recycling costs are recovered. 

Among the discussed solutions, solution 1 can be implemented easily. The complexity of Solution 2 is higher than Solution 1, but its implementation is not too challenging. We can use any platform that supports smart contracts. In solution 2, we also have to rely on existing digital currencies and use stablecoins backed by fiat currencies to minimize the price variation risks. Finally, solution 3 is challenging and requires bootstrapping and the development of appropriate marketplaces for coin trading. A lot of consistent effort is generally needed to ensure the success of any new digital coin. As DeFi protocols mature in the future, we believe that the technical challenges related to price fluctuations can be resolved. However, we argue that the introduction of native coins in the system will make smart contract execution more efficient by making them an integral part of the decentralized application (DApp) ecosystem. Furthermore, this opens up the possibility to popularise the initiative and attract more participation like the Powerledger initiative in Australia \cite{Mike_Butcher_2018} \cite{EY_Oceania_2023}.

Moreover, given the significance of this problem and by building the interest of solar manufacturers, regulators, and recycling companies, prosumer participation, marketplace creation, and other economic challenges could be resolved. Complexity levels in these three scenarios are different. We argue that someone should start with the relatively less complex system presented in scenario 1. The consortium can expand to scenarios 2 and 3 based on the user feedback and operating environment. Furthermore,  In the blockchain, there is no central body that is trusted by everyone. Hence, a distributed ledger like blockchain will have the benefit of incorporating companies from different countries. Indeed, there could be issues where certain companies may have specific regularisation due to country limitations. In such cases, it is expected that the system will comply with the local regulations of the country where it is being used. Examples of such compliance with local regulations can be found in other multi-national companies such as Google and Facebook which operate in many different countries within their own regulatory frameworks.

Finally, energy consumption is a big concern for blockchain-based systems. However, private and consortium blockchain energy consumption is relatively much lower compared to public blockchain. According to a study \cite{sedlmeir2020energy}, a simple key-value store such as LevelDB can sustainably operate tens of thousands of transactions per second with a power consumption of less than 100 W, which yields less than 10\textsuperscript{-2} J per transaction. A more complex database, such as CouchDB, with one backup still manages more than 10\textsuperscript{3} transactions per second on the same hardware, resulting in at most 0.1 J per transaction. As an example of a small-scale enterprise blockchain, such as Hyperledger Fabric architecture with 10 nodes, each on cloud instances with 32 vCPUs and therefore likely consuming a few thousand Watts in total. Such a system can handle around 3000 transactions per second, so we arrive at an order of magnitude of 1 J per transaction. Finally, public blockchain such as Bitcoin consumes about 10\textsuperscript{9} J per transaction. Considering the energy consumption of different blockchain types, our choice of consortium blockchain is more justified.

\begin{table}[t]
\caption{Summary of the three proposed solutions.}
\begin{tabular}{|p{2cm}|p{2cm}|p{3cm}|p{4.5cm}|p{3cm}|}
\hline
\textbf{Solution}   & \textbf{Currency}          & \textbf{Probability of financial gain  }                                         & \textbf{Use of blockchain }   & \textbf{ Innovation }                  \\ \hline

Solution 1 & Fiat currency     & N/A                                                                           &  Data storage and management   & Scope is comparatively limited       \\ \hline
Solution 2 & Stablecoin (e.g, CBDC)  & Moderate & In addition to data management, it also facilitates digital currency transfers and eliminates traditional banks from the loop & Enables innovative services
\\ \hline
Solution 3 & Traditional cryptocurrency & High (depends on the popularity of the platform).  & Blockchain has its native currency and the policies are managed and enforced by smart contacts  & Facilitate more innovative services e.g., staking. \\ 
\hline
\end{tabular}
\label{Table:compare}
\end{table}

In Table~\ref{Table:compare}, we show a summary and comparison of our three proposed solutions. Solution 1 is mainly focused on describing the main idea of co-sharing responsibilities of recycling among the stakeholders. Solution 2 brings the concept of stable coin in addition to data management which makes the system more seamless compared to solution 1. It can also support more complex smart contract functionalities. In addition to solutions 1 and 2, solution 3 provides traditional native crypto coins which may help to acquire early adopters and raise funds for the proposed systems. One relevant example could be PowerLedger, which is traded in crypto exchange and current market cap of 180 million USD~\cite{Powerledger2024}.

\section{Conclusion}
Although the importance of solar panels in achieving a future net zero is paramount, the lack of initiatives on how to recycle these panels after their end-of-life is putting the future environment in jeopardy. The cost of recycling is very high and as a result, recycling companies - the solely responsible entities for recycling solar panels - struggle to find a profitable business case. Because the world will create an enormous amount of solar e-waste by 2050, this paper has taken the first step to propose a division of the recycling cost among all the relevant stakeholders - beneficiaries of using clean solar energy.  Three blockchain-based solutions have been proposed that stimulate responsibility sharing through the circular use of solar energy produced by the panels and keep track of solar panels' end-of-life and stakeholders' responsibilities for circular recovery. By tracking the energy-related activities of the users, the degradation of panels can be monetized for environmental recycling. According to the proposed solutions, prosumers pay the recycling fees gradually over the long lifespan of the solar panels, while other stakeholders contribute by maintaining the blockchain infrastructure. Among the proposed solutions, although a solution that uses blockchain for information and data storage is easy to implement, the solution that has been proposed using a recycling coin (RC-coin) to incentivize solar panel recycling is more effective for achieving the full potential of blockchain and digital currencies. Our system is adaptable and can conform to various regulatory requirements. For instance, it can be tailored to accommodate an eco-tax, an extra charge applied to the selling price of photovoltaic systems. This is achieved by configuring the consortium blockchain system to align with the specific operational context.

A prospective expansion of this research involves conducting a thorough analysis using real data, engaging with diverse stakeholders, and presenting the advantages of the proposed approach through a more comprehensive methodological framework. Another intriguing avenue for future research involves adapting the mechanism outlined in the paper to facilitate the establishment of collaborative responsibility in the realm of energy storage recycling. Additionally, we aim to explore different regulatory and financial factors in the future by engaging with industries and policymakers working to reform the waste management landscape of the future.

\section*{Acknowledgement}
This research was supported in part by LUMS Faculty Initiative Fund (FIF) and in part by the National Research Foundation, Singapore, and Infocomm Media Development Authority under its Future Communications Research \& Development Programme, DSO National Laboratories under the AI Singapore Programme (AISG Award No: AISG2-RP-2020-019 and FCP-ASTAR-TG-2022-003), and MOE Tier 1 (RG87/22). The work of H. V. Poor was supported in part by a grant from the Princeton University School of Engineering and Applied Science.

\section*{Biographies}
\textbf{Dr. M. J. M. Chowdhury} (m.chowdhury@latrobe.edu.au) is currently working as a Senior Lecturer in Cybersecurity at La Trobe University, Australia. He has received the PhD degree from Swinburne University of Technology, Melbourne, Australia. He is currently working with blockchain, privacy, and Trust. He has published his research in top conferences and journals in computer science. He also worked as a TPC member in different top-tier conferences.

\textbf{Naveed Ul Hassan} (naveed.hassan@lums.edu.pk), an Associate Professor of Electrical Engineering at Lahore University of Management Sciences (LUMS), earned his B.E. from Risalpur, Pakistan, in 2002, and the M.S. and Ph.D. from Ecole Superieure d’Electricite, France, in 2006 and 2010. With a robust academic background, he has published 100+ research papers in international journals and conferences. His diverse research spans blockchain, wireless communication, smart energy systems, and indoor positioning. Naveed has played a pivotal role in establishing a blockchain technologies academic program at LUMS and serves as an Associate Editor for Computer Science (Springer Nature) and IET Smart Grid.

\textbf{Wayes Tushar} (w.tushar@uq.edu.au) obtained his B.Sc. in electrical and electronic engineering from Bangladesh University of Engineering and Technology in 2007, followed by a Ph.D. in engineering from The Australian National University (ANU) in 2013. Presently, he is a Senior Lecturer at the School of Electrical Engineering and Computer Science (EECS), The University of Queensland (UQ). His research focus revolves around energy management, especially in the area of energy sharing and trading.

\textbf{Dusit Niyato} (dniyato@ntu.edu.sg) is a professor in the College of Computing and Data Science, at Nanyang Technological University, Singapore. He received Ph.D. in Electrical and Computer Engineering from the University of Manitoba, Canada in 2008. His research interests are in the areas of sustainability, edge intelligence, decentralized machine learning, and incentive mechanism design. Dusit’s research experience includes the use of optimization and machine learning techniques to solve various resource allocation issues in these topics for more than 15 years.

\textbf{Tapan K. Saha} (saha@eecs.uq.edu.au) is a Professor of Electrical Engineering at The University of Queensland (UQ). He earned his B.Sc.Eng. from Bangladesh University of Engineering and Technology in 1982, M.Tech. from Indian Institute of Technology Delhi in 1985, and Ph.D. from UQ in 1994. Currently, he leads the UQ Solar and Industry 4.0 UQ Energy TestLab and serves as the Founding Director of the Australasian Transformer Innovation Centre. His expertise lies in condition monitoring of electrical assets, power systems, and renewable energy integration. A Fellow of IEEE  and a Fellow \& Chartered Professional Engineer of the Institution of Engineers, Australia, he is also a Registered Professional Engineer in the State of Queensland.

\textbf{H. Vincent Poor} (poor@princeton.edu) is the Michael Henry Strater University Professor at Princeton University, where his interests include information theory, machine learning and network science, and their applications in wireless networks, energy systems, and related fields. He is a member of the U.S. National Academy of Engineering and the U.S. National Academy of Sciences, and a foreign member of the Royal Society and other national and international academies. Among his publications is the book \emph{Advanced Data Analytics for Power Systems} (Cambridge, 2021).

\textbf{Chau Yuen} (chau.yuen@ntu.edu.sg) received the B.Eng. and Ph.D. degrees from Nanyang Technological University, Singapore, in 2000 and 2004, respectively. Since 2023, he has been with the School of Electrical and Electronic Engineering, Nanyang Technological University. Dr. Yuen received the IEEE Communications Society Fred W. Ellersick Prize (2023), the IEEE Marconi Prize Paper Award in Wireless Communications (2021), and several other best paper awards. He received the IEEE Asia Pacific Outstanding Young Researcher Award in 2012. He is a Highly Cited Researcher by Clarivate Web of Science. He has been working on energy management-related projects, especially security in cyber-physical systems. 


\end{document}